\documentclass[conference,a4paper]{IEEEtran}
\usepackage{graphicx}
\usepackage{epstopdf}
\usepackage{cite}
\usepackage{amsfonts}
\usepackage{amssymb}
\usepackage{amsmath}
\usepackage{mathrsfs}
\usepackage{bm}
\usepackage{amsmath}
\usepackage{mdwmath}
\usepackage{mdwtab}
\usepackage[ruled,vlined,linesnumbered]{algorithm2e}

\begin{document}

\title{Elastic Local Breakout Strategy and Implementation for Delay-Sensitive Packets with Local Significance}
\author{\IEEEauthorblockN{Yongkang Li, Zhiyuan Jiang, An Xu, Sheng Zhou, and Zhisheng Niu, \emph{Fellow, IEEE}}
   \IEEEauthorblockA{Tsinghua National Laboratory for Information Science and Technology,\\
        Department of Electronic Engineering, Tsinghua University, Beijing, 100084, China\\
        liyk15@mails.tsinghua.edu.cn, zhiyuan@tsinghua.edu.cn, anxu@mails.tsinghua.edu.cn, \\
        \{sheng.zhou, niuzhs\}@tsinghua.edu.cn}}

\maketitle 

% As a general rule, do not put math, special symbols or citations
% in the abstract
\begin{abstract}
Explosion of mobile traffic will bring a heavy burden to the core network, and rapid growth of mobile devices, as well as increasing demand for delay-sensitive services poses severe challenges to future wireless communication systems. In this regard, local breakout is a promising solution to save core network load and, more importantly, to reduce end-to-end (e2e) delay of packets with local significance. However, the capacity of local breakout link is limited, resulting in excessive delay when the traffic load through the local link is high. Therefore, the decision on whether the traffic flows should be transmitted through core network or by local breakout link has great practical significance. In this paper, we propose and implement a novel local breakout framework to deliver low e2e delay packets with local significance. A real-time local breakout rule based on the solution to a Markov decision process is given, showing that some packets with local significance should pass through core network rather than being delivered by local breakout link to meet the delay requirements. To test our proposed framework, a long-term-evolution (LTE) based test-bed with virtual base stations is implemented, by which we show the proposed framework is feasible and the e2e delay is significantly reduced.
\end{abstract}

\section{Introduction}
\label{section1}
With the flourishing of mobile internet services, and the emergence of machine-type communications, mobile data traffic increased dramatically in recent years. Cisco forecasts that overall mobile data traffic is expected to  be a seven fold increase over 2016 by 2021, and mobile devices and connections will grow to 11.6 billion by 2021 \cite{index520862global}, which will cause heavy load on core network because all packets should pass through core network in accordance with the 3rd Generation Partner Project (3GPP) standards. Moreover, $5$G communication systems not only provide high data rate services \cite{Andrews14}, but also promise to deliver packets with very low e2e latency for delay-sensitive applications \cite{boc14}, e.g., vehicle-to-everything (V2X) communications, machine-type communications and etc. $5$G systems allow information to be transmitted with stringent delay requirements (milliseconds) and low drop ratio, and thus enable various promising vertical applications for $5$G and beyond.

One important feature of most packets with stringent delay requirements is that the information often exhibits \emph{local significance} \cite{zheng15}, which indicates that the delay and quality-of-service (QoS) requirements are higher for packets whose transmitter and receiver are geographically nearer. For instance, the information in vehicular networks often has local significance. The safety information, e.g., breaking and lane-changing alerts, usually requires much lower delay and better QoS, compared with navigation and infotainment system information. Meanwhile, the necessary spread area for safety information is lower than, e.g., navigation information. Similar consequences also apply in, e.g., status-monitoring systems and smart grid systems where a nearby failure must be known more quickly to make appropriate adaptation compared with a remote one. Essentially, it is common sense that people are more concerned to acquire information about the things that happen near them than those far from them, because the influence of the local information is far more likely to manifest momentarily. In addition, Caching popular contents in base stations may be a trend in future\cite{wang2014cache}. To share cache contents between adjacent cache servers is one way to solve the limited cache capacity issue in this regard.

Local breakout is a technique which has drawn increasing attention to solve this problem. It essentially bypasses the core network for local packets to reduce the e2e delay. Local Internet Protocol (IP) Access (LIPA) and Selected IP Traffic Offload (SIPTO) have already been defined in the the $3$rd Generation Partnership Project (3GPP) standards to realize local breakout \cite{local2011access}. LIPA and SIPTO enable operators to offload certain types of traffic to a home-network or a pre-defined IP network at a network node close to the user equipment. However, LIPA/SIPTO only defined the architecture and procedures of breakout, there is no specific strategy on how to realize local breakout and choose breakout path based on network status.
 
A Converged Gateway (CGW) which has local SIPTO function, is introduced in \cite{cartmell2013local}. Whenever an uplink-initiated traffic flow from the Wi-Fi access point or base station starts, the CGW will select an application server to serve the end-user device who is establishing the connection, bypassing the core network. However, the framework should introduce a new CGW unit, which will result in higher deployment cost. In \cite{yang2013solving}, the authors propose that adjacent UEs can set up a device-to-device (D2D) link using a cellular interface, and the traffic packets can be directly exchanged through the D2D link. However, this solution will cause excessive energy consumption of UEs, and the communication distance is limited. The authors introduce Mobile Edge Computing (MEC) into the local breakout framework in  \cite{zhang2016mobile}\cite{lee2016local}, and the MEC unit will be deployed between base stations and core network. However, this framework requires all IP flows passing through MEC unit, which may cause failure of the single point, and it will take a long time to fully implement this cloud-based architecture in future. What's more, in the study above, there are few researches on the local breakout strategy.
  
There are some other efforts toward optimizing the strategy for local breakout. It is proposed that one can differentiate types of services based on different Public Data Network (PDN) connections \cite{samdanis2012traffic} or different bearers \cite{ma2011traffic}. In \cite{katanekwa2013mobile}, the authors focus on optimizing the path of local breakout. In \cite{lee2014novel}, a local breakout strategy is formulated according to network utilization. However, all above researches only give preliminary strategies, there is no analysis about whether the packets delivered by local breakout can meet delay requirement, the path choice problem of packets in accordance with the requirements of e2e latency remains open. 
 
In this paper, we propose a new local breakout framework to deliver low e2e delay packets between neighbouring base stations with local significance, and we give a Markov-decision-process based optimum scheduling strategy solved by value iterations based on the Bellman equations \cite{bert95}, showing that some packets with local significance should pass through core network rather than being delivered by local breakout to meet the delay requirements. To test the feasibility of our proposed local breakout scheme, we implement a software-defined-radio platform based on Open Air Interface (OAI) \cite{nikaein2014openairinterface}. Measurements show that the e2e delay is significantly reduced by the proposed local breakout scheme.

\section{Local Breakout Between Adjacent Base Stations }
\label{section2}

\subsection{Local Breakout Framework}
\begin{figure*}[!t]
    \centering
    \includegraphics[width=0.9\textwidth]{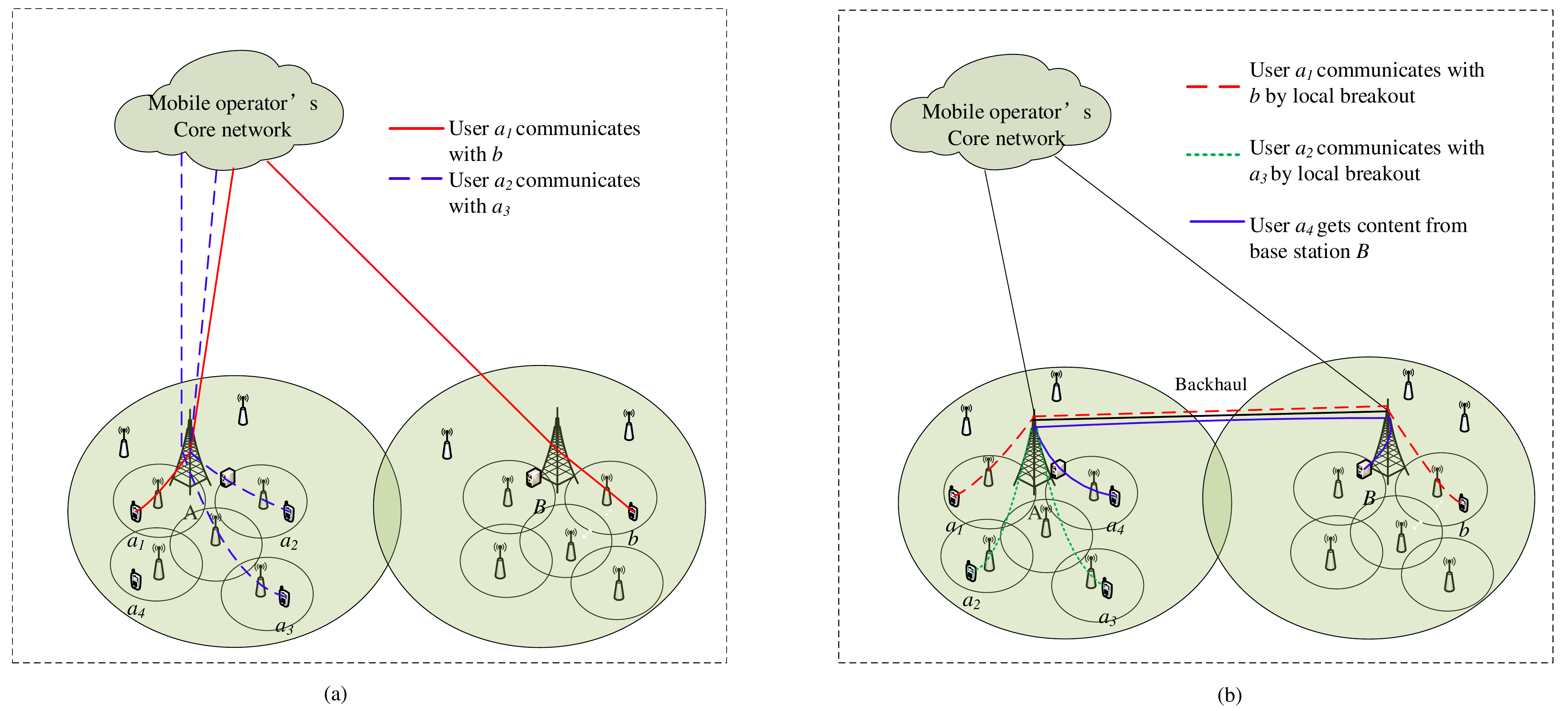}
    \caption{The high-level framework of the existing network and our proposed local breakout network. (a) shows a sketch of an ultra-dense distributed cache network. (b) shows a sketch of our new network framework which introduces the direct link between adjacent base stations and local breakout functionality.
    }
    \label{cache_frame_lb}
\end{figure*}

When applying the local breakout concept into practice, we consider a cellular network with a \emph{Control base station} (CBS) and several \emph{Traffic base stations} (TBSs).
Fig. \ref{cache_frame_lb} shows the architecture of the existing network and our proposed local breakout network. Fig. \ref{cache_frame_lb} (a) shows the sketch of an ultra-dense content cache network, where one CBS is responsible for controlling several TBSs within its coverage. Considering the following scenario, when users wants to fetch some files which are not cached in base station $A$, it must fetch files from network with all IP packets passing through core network even if the files are cached in adjacent base station $B$, because adjacent base stations can't share cache contents through the backhaul in this architecture. What's more, as there is no local breakout, service traffic flows also need to pass through core network when user $a_2$ wants to communicate with user $a_3$ or user $a_1$ wants to communicate user $b$, just as shown in the figure.

Correspondingly, Fig. \ref{cache_frame_lb} (b) shows a sketch of our new framework with local breakout implemented. In this framework, files cached in adjacent base stations can be shared through the backhaul, thus users $a_4$ can fetch some files from base station $B$ directly, and the service traffic flows can also be transmitted directly within a single base station or through the link between adjacent base stations, without passing through the core network, as shown in the figure.
\subsection{Local Breakout Protocol }
\begin{figure}[!t]
    \centering
    \includegraphics[width=0.45\textwidth]{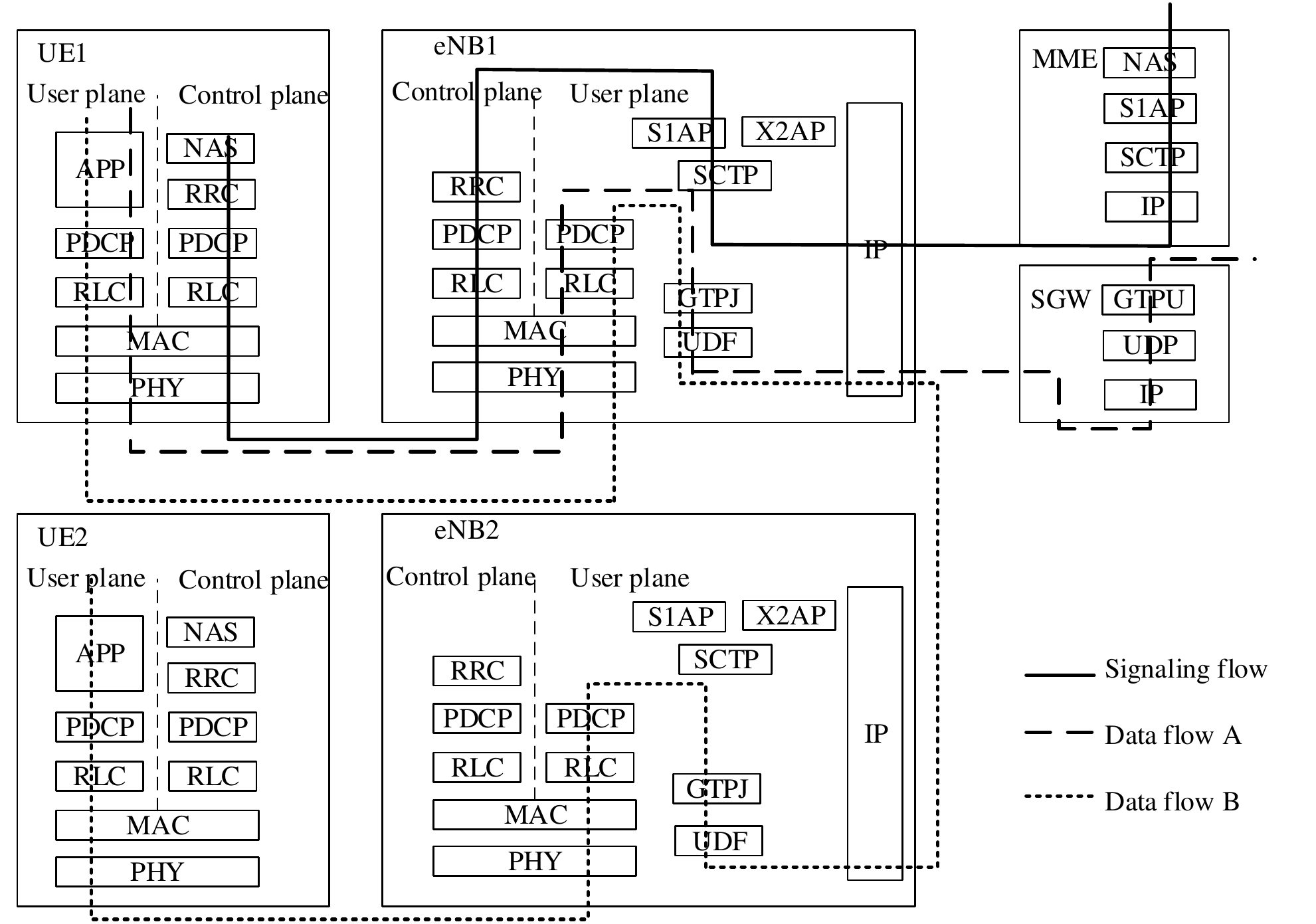}
    \caption{A novel protocol for local breakout.}
    \label{Fig_sys_protocol_lb}
\end{figure}

The LTE protocol structure can be divided into data plane and control plane. According to the 3GPP standards, all IP flows,  whatever data flows or signaling flows, should pass through core network by General Packet Radio Service (GPRS) Tunnelling Protocol (GTP), as data flow $A$ and signaling flow shown in Fig. \ref{Fig_sys_protocol_lb}, which entail heavy load to the core network.

Compared with LTE protocol, our proposed new protocol shows that the data flows can be directly delivered from UE$1$ to UE$2$ by local breakout along the link of data flow $B$, bypassing core network, and signaling flows still pass through core network as before. 

The main idea of our local breakout framework is directly delivering IP packets with local significance between neighbouring base stations, bypassing core network. The base station can obtain all IP packets from the user by decapsulation of GTP protocol, and hence it can change the path to adjacent base station which are originally scheduled to be forwarded to the core network based on a breakout rule for IP packets, which can reduce the e2e delay of packets to a extent, and it will be validated by real measurements in Section VII.

\begin{figure}[!t]
    \centering
    \includegraphics[width=0.35\textwidth]{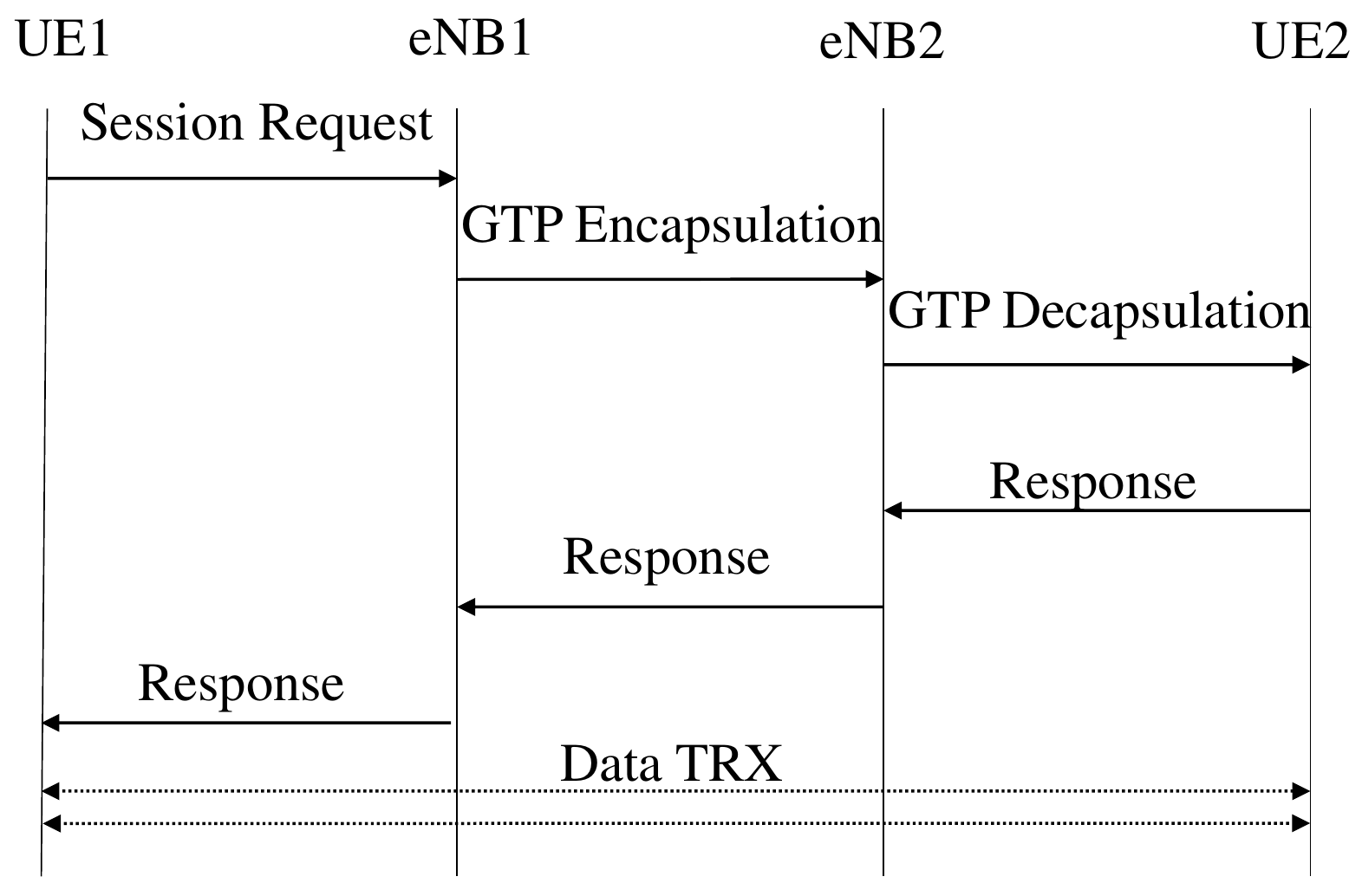}
    \caption{Information flows between mobile users and the base stations.}
    \label{Fig_sys_inform}
\end{figure}

In our local breakout framework, the CBS should have domain name system (DNS) resolution function and the ability to realize traffic flows breakout based on specified IP address.  Fig. \ref{Fig_sys_inform} shows the information flows between mobile users and base stations based on the local breakout framework. For instance, UE-$1$ wants to communicate with UE-$2$, and when the base station receives the request from UE-$1$, it will encapsulate request information and find, based on DNS resolution function and the signaling information, the target user UE-$2$ is attached to its adjacent base station,  then eNB-$1$ will deliver the packets into adjacent base station eNB-$2$ based on specified target IP address. And eNB-$2$ will then send a response, and then UE-$1$ and UE-$2$ can transmit IP packets directly after the channel is built.
\section{System Model}
\label{section3}
In our proposed framework of local breakout, as the bandwidth of backhaul between adjacent base stations is limited \cite{jiang_icc17}, and part of the bandwidth needs to be reserved for dedicated services, the link transmission capacity for local breakout may become a bottleneck, resulting in excessive delay when the traffic load through the local link is high. Therefore, we need study the decision strategy of local breakout. We try to give a real-time decision about whether the traffic flows should be transmitted through the core network or through the breakout link according to the demand of traffic flows latency and the network status.

\subsection{Tasks}
\begin{figure}[!t]
    \centering
    \includegraphics[width=0.45\textwidth]{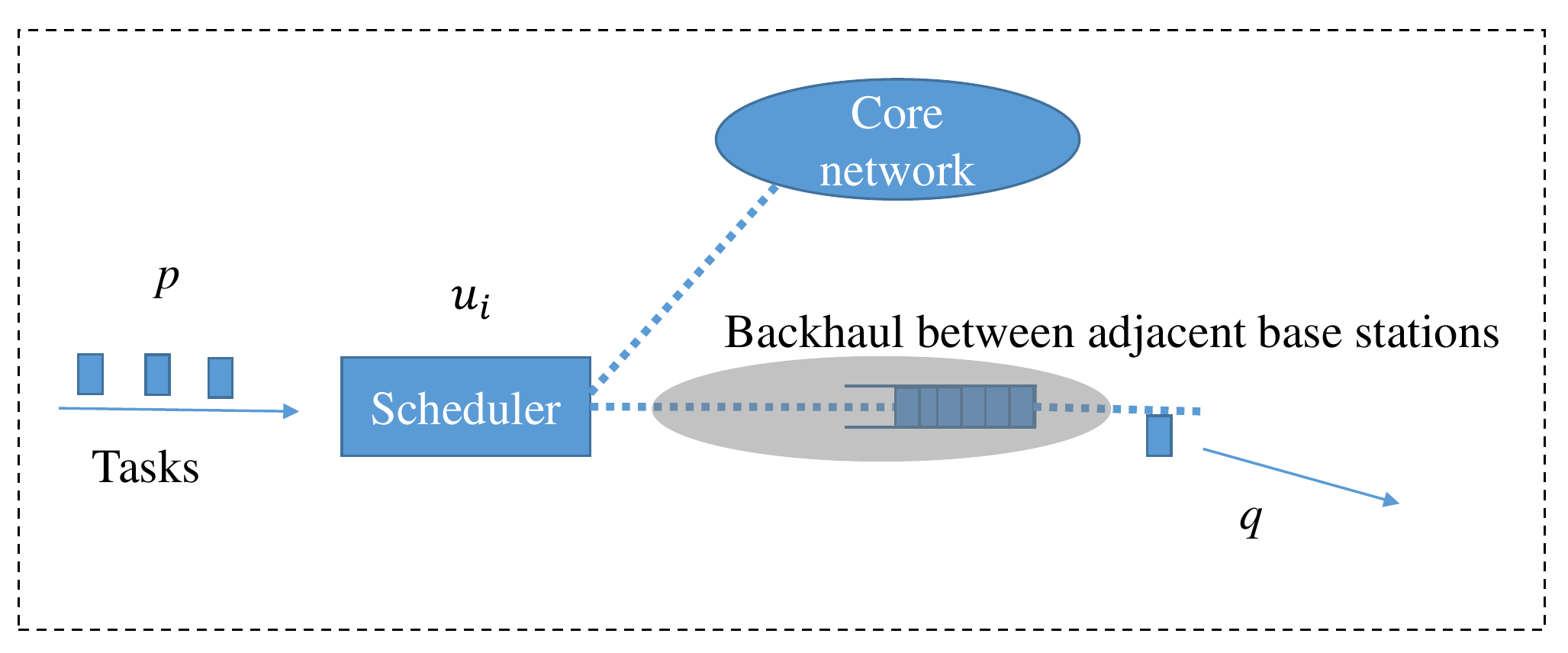}
    \caption{System model}
    \label{Fig_sys_sys_model}
\end{figure}

The system model is shown in Fig. \ref{Fig_sys_sys_model}. The traffic flows are firstly transmitted to a scheduler which is located in the CBS. Each uplink-initiated traffic flow from mobile devices is considered as an arriving packet, and the scheduler decides whether to transmit the packets by local breakout or send them to the core network. A packet will queue in line if there are many packets in front of it when it is transmitted to the brreakout backhaul. We consider that the packet is successfully completed if it is transmitted within its delay constraint. Our goal is to design a policy that assigns the packets to core network or to be delivered within a single base station or adjacent base stations by local breakout in an online fashion to maximize the success probability.
The number of arrival packets and the number of departure packets from the backhaul, in each time slot $\tau$, are random variables which are both bernoulli distributed with arrival probability $p$ and departure probability $q$ respectively. And in each time slot, there is only one arrival or departure of packet at most.

\subsection{Markov Decision Process Based Formulation}
We formulate the above-mentioned problem as an Markov Decision Process (MDP) \cite{4}. An assignment policy needs to make decision only at the time when a packet arrives at the scheduler.

The system state when the $i$-th packet arrives at the scheduler is described by $Q_{i}$, where $Q_{i}$ denotes the length of the queue in the backhaul between adjacent base stations. The scheduler's decision is denoted as $u_{Q_{i}}$ , where $u_{Q_{i}}$ equals to $1$ if the arrival packet is assigned to the backhaul link, and $u_{Q_{i}}$ equals to $1$ if the arriving packet is assigned to the core network. The system delay constraint of the arriving packet is $T_{i}\tau$.
Thus, the state of the system evolves as, when $u_{Q_{i}}=1$,
\begin{IEEEeqnarray}{rCl}
  &&P\{Q_{i+1}|Q_{i}\}=\nonumber \\
  &&\begin{cases}
  \sum_{n=1}^{\infty}p(1-p)^{n-1}\binom{n}{Q_{i}+1-Q_{i+1}}  \\
   \ \  \cdot (1-q)^{(n-(Q_{i}+1-Q_{i+1}))} \\
   \ \  \cdot q^{(Q_{i}+1-Q_{i+1})}, & \mbox{$Q_{i+1}\leq Q_{i}+1$};\\
  0, &\mbox{otherwise}.
  \end{cases}
\end{IEEEeqnarray}
When $u_{Q_{i}}=0$,
\begin{IEEEeqnarray}{rCl}
  &&P\{Q_{i+1}|Q_{i}\}=\nonumber \\
  &&\begin{cases} 
  \sum_{n=1}^{\infty}p(1-p)^{n-1}\binom{n}{Q_{i}-Q_{i+1}}  \\
  \ \  \cdot (1-q)^{(n-(Q_{i}-Q_{i+1}))}q^{(Q_{i}-Q_{i+1})}, &
  \mbox{$Q_{i+1}\leq Q_{i}$};\\
  0, &\mbox{otherwise}.
  \end{cases}
\end{IEEEeqnarray}
where $n$ denotes the interval time is the $n$ time slots between adjacent arrival packets, and $P\{Q_{i+1}|Q_{i}\}$ is the transition probability of system from state $Q_{i}$ to state $Q_{i+1}$.

We assume that if a packet is successfully transmitted, a reward of $A$ is given, and otherwise the reward is $0$. Then our goal is to design a packet assignment policy that maximizes the system reward. The average reward incurred when a packet is scheduled is given by
\begin{equation}
\begin{aligned}
 \frac{1}{K}E\left[\sum_{i=1}^{K}(u_{Q_{i}}R_\textrm{bac}(Q_{i})+(1-u_{Q_{i}})R_\textrm{cn}(Q_{i}))\right],
\end{aligned}
\end{equation}
where,
\begin{equation}
\begin{aligned}
  R_\textrm{bac}(Q_{i})=P_{\textrm{bac},i}(t<T_{i}) A,
\end{aligned}
\end{equation}
\begin{equation}
\begin{aligned}
 R_\textrm{cn}(Q_{i})=P_{\textrm{cn},i}(t<T_{i}) A.
\end{aligned}
\end{equation}
where $K$ is the number of decisions, $E[\cdot]$ is the expectation, $P_{\textrm{bac},i}(t<T_{i})$ and $P_{\textrm{cn},i}(t<T_{i})$ are the probability that $i$-th packet is successfully transmitted through backhaul between adjacent base stations and core network respectively.
It follows that
\begin{equation}
\begin{aligned}
 P_\textrm{bac}(i)=1-\sum_{j=1}^{Q_i}\binom{j}{T_i}q^j(1-q)^{(T_i-j)}.
\end{aligned}
\end{equation}
where $j$ is the total packets departing from backhaul within the interval between adjacent arrival packets.
$P_\textrm{cn}(i)$ can be obtained based on the Internet transmission delay distribution. Some related works model Internet transmission delay \cite{hooghiemstra2017delay} by a density  $\varphi(t)$ of the following form
\begin{equation}
\begin{aligned}
\varphi(t)=\alpha \varphi_2(t)+(1-\alpha)\varphi_{1}(t)*\varphi_{2}(t) , \ \ t\geq 0.
\end{aligned}
\end{equation}
where $\varphi_{1}(t)$ is the Internet traffic delay which is obey exponential distribution and $\varphi_{2}(t)$ is the router processing delay which is obey Gaussian distribution.

Because the queue length $Q_{i}$ can change in any time after a packet is assigned, the problem can be transformed as 
\begin{equation}
\begin{aligned}
  \max \limits_{\pi}  \lim\limits_{K\to\infty}\inf \frac{1}{K}E\left[\sum_{i=1}^{K}(u_{Q_{i}}R_\textrm{bac}(Q_{i})+(1-u_{Q_{i}})R_\textrm{cn}(Q_{i}))\right],
\end{aligned}
\end{equation}
\begin{equation}
\begin{aligned}
  s.t. \ \  u_{Q_{i}}\in {\{0,1\}}.
\end{aligned}
\end{equation}
where $\pi$ is any causal (i.e., history independent) policy.

Let $V_{K}(s)$ denotes the $K$-horizon optimal cost-to-go incurred by the system starting in initial state $Q_{{0}}=s$. Then, the optimal cost-to-go function satisfies the following recursive formula,
\begin{equation}
\begin{aligned}
  V_K(s)=\max \limits_{u_s\in {\{0,1\}}} & \left\{u_s R_\textrm{bac}(s)+(1-u_s)R_\textrm{cn}(s)\right. \\
    & \left. +\sum_{\bar{s}=0}^{s+u_s}[\sum_{n=s+u_s-\bar{s}}^{\infty}(1-p)^{(n-1)p}\binom{n}{s+u_s-\bar{s}} \right. \\
  & \left. (1-q)^{n-(s+u_s-{\bar{s}})}q^{(s+u_s-\bar{s})}]V_{K-1}(\bar{s})\right\}.
\end{aligned}
\end{equation}
where $\bar{s}$ is the system state subsequent to state $s$ when a new packet arrives and is assigned to backhaul or core network.

\subsection{Optimal Stationary Policy by Value Iteration}
The above-mentioned problem is a infinite horizon average problem, to obtain a stationary optimal policy, the queue length needs to be truncated to make it a finite-state system. we can calculate with the following algorithm:  
 We firstly let $V_{0}(s)=0$, and $V_{K}(s)$ can be determined by $V_{K-1}(s)$ with the recursive relationship in (10). and then we let $V_{K}(s)=V_{K}(s)-V_{K}(0)$, $\forall s$. This process is continued until the optimal relative optimal cost to-go functions remain the same after improvement, i.e.  $|V_{K}(s)-V_{K-1}(s)|/|V_{K-1}(s)|<\epsilon$ where $\epsilon$ is a small positive number. Then, we can get the optimal policy, maximizing (8), that assigning the arriving packet to the backhaul or core network when the system state is $s$.

\section{Simulations}
\label{section4}
We present the results of a simulation study for local breakout scenario. The parameters are set as follows: $\tau=0.02$~ms, $T_{i}\tau=19.3$~ms; the buffer size of backhaul is $100$. The internet delay is modeled in (7), whose mean delay is $40$~ms.

\begin{figure}[!t]
    \centering
    \includegraphics[width=0.35\textwidth]{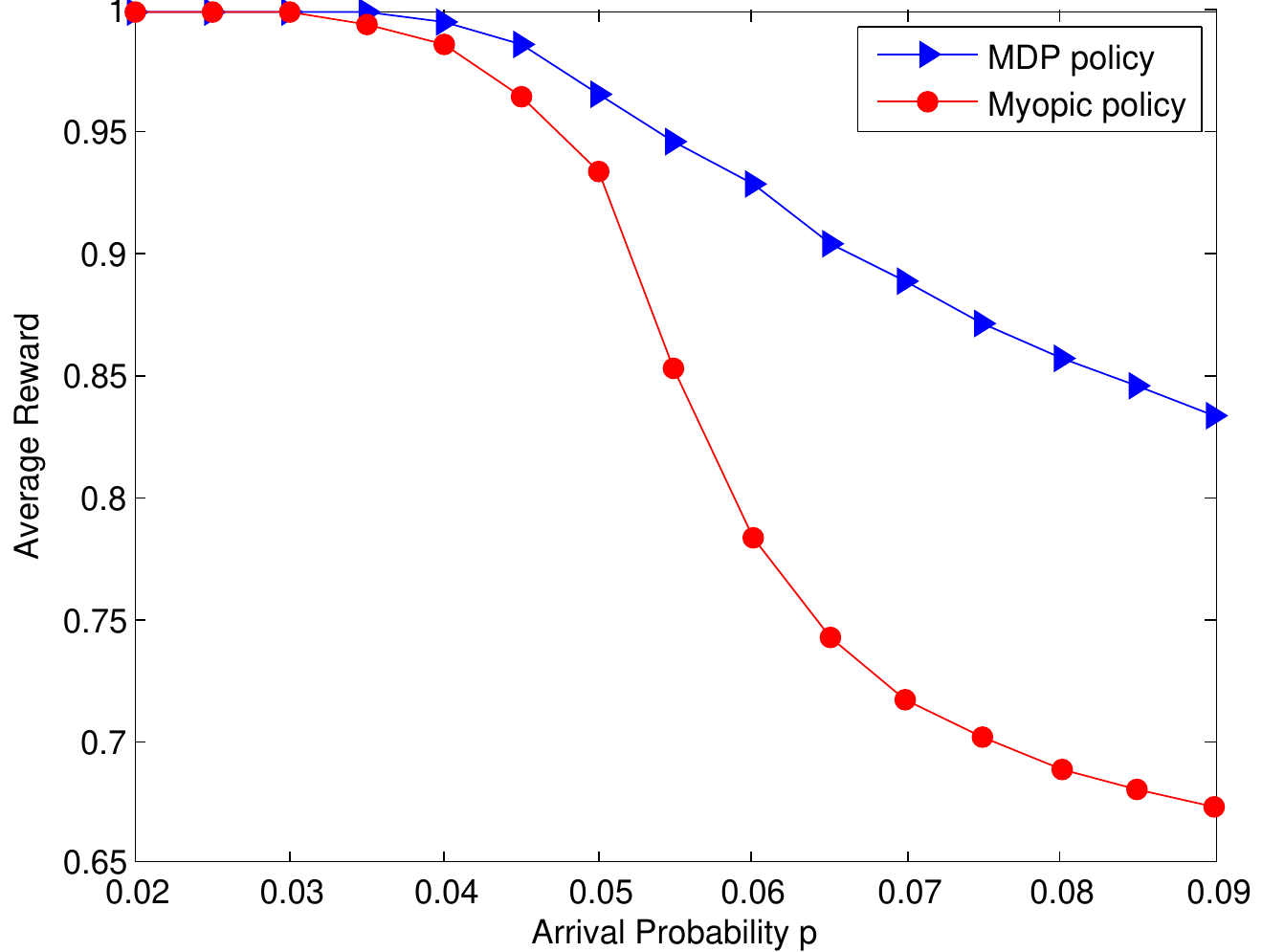}
    \caption{Average reward vs. arrival probability for MDP policy and myopic policy.}
    \label{Fig_sys_Ave_Reward_p}
\end{figure}

\begin{figure}[!t]
    \centering
    \includegraphics[width=0.35\textwidth]{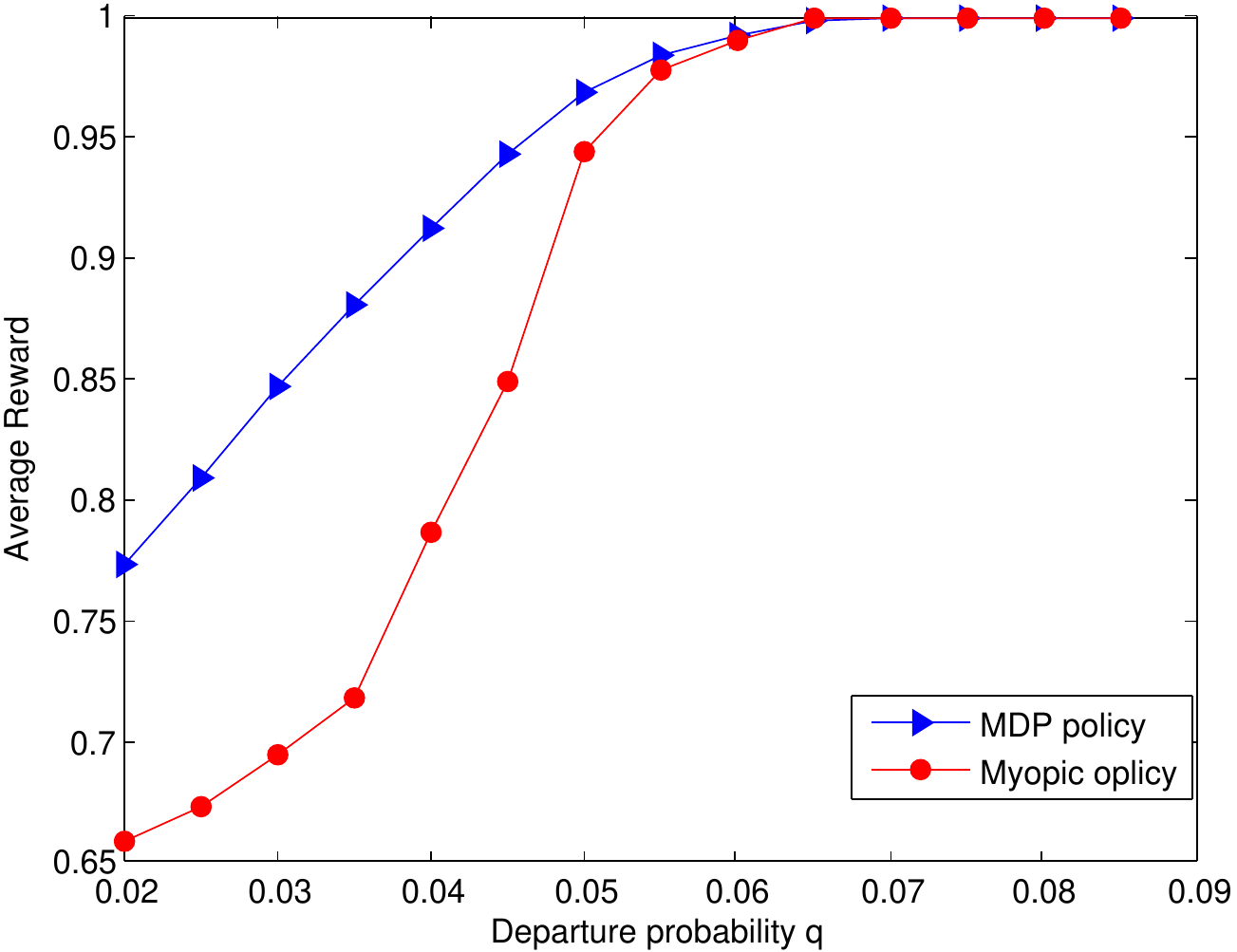}
    \caption{Average reward vs. departure probability for MDP policy and myopic policy.}
    \label{Fig_sys_Ave_Reward_q}
\end{figure}

Fig. \ref{Fig_sys_Ave_Reward_p} and Fig. \ref{Fig_sys_Ave_Reward_q} present the performance of both the MDP policy and the myopic policy. The myopic policy is defined as the policy that maximizes the immediate reward (one-step reward). These two policies are compared according to their normalized average system reward for different arrival probability $p$ and different departure probability $q$. In Fig. \ref{Fig_sys_Ave_Reward_p}, departure probability holds constant value of $q=0.05$. It can be seen that the difference of average reward of two policies are obvious, the relative difference is about $30\%$ when arrival probability $p$ equals $0.09$. Moreover, we can see that when the arrival probability is low, the average rewards, in terms of the success probability, of both policies are high. However, the average reward decreases with the increase of arrival probability, which can be interpreted as follows: most packets in the backhaul have to wait in the queue when traffic load is high, which leads to poor QoS and then further leads to low success probability. On the other hand, given a constant arrival probability $p=0.05$, we compare the performance of these two policies with different departure probability. Fig. \ref{Fig_sys_Ave_Reward_q} shows that in our model, MDP policy results in an approximately $20\%$ average reward improvement when departure probability $q$ equals $0.02$, compared with myopic policy. And with the increasing of departure probability, the average reward also increases, because the higher departure probability means the lower traffic load in the backhaul, leading to a higher success probability.

\begin{figure}[!t]
    \centering
    \includegraphics[width=0.35\textwidth]{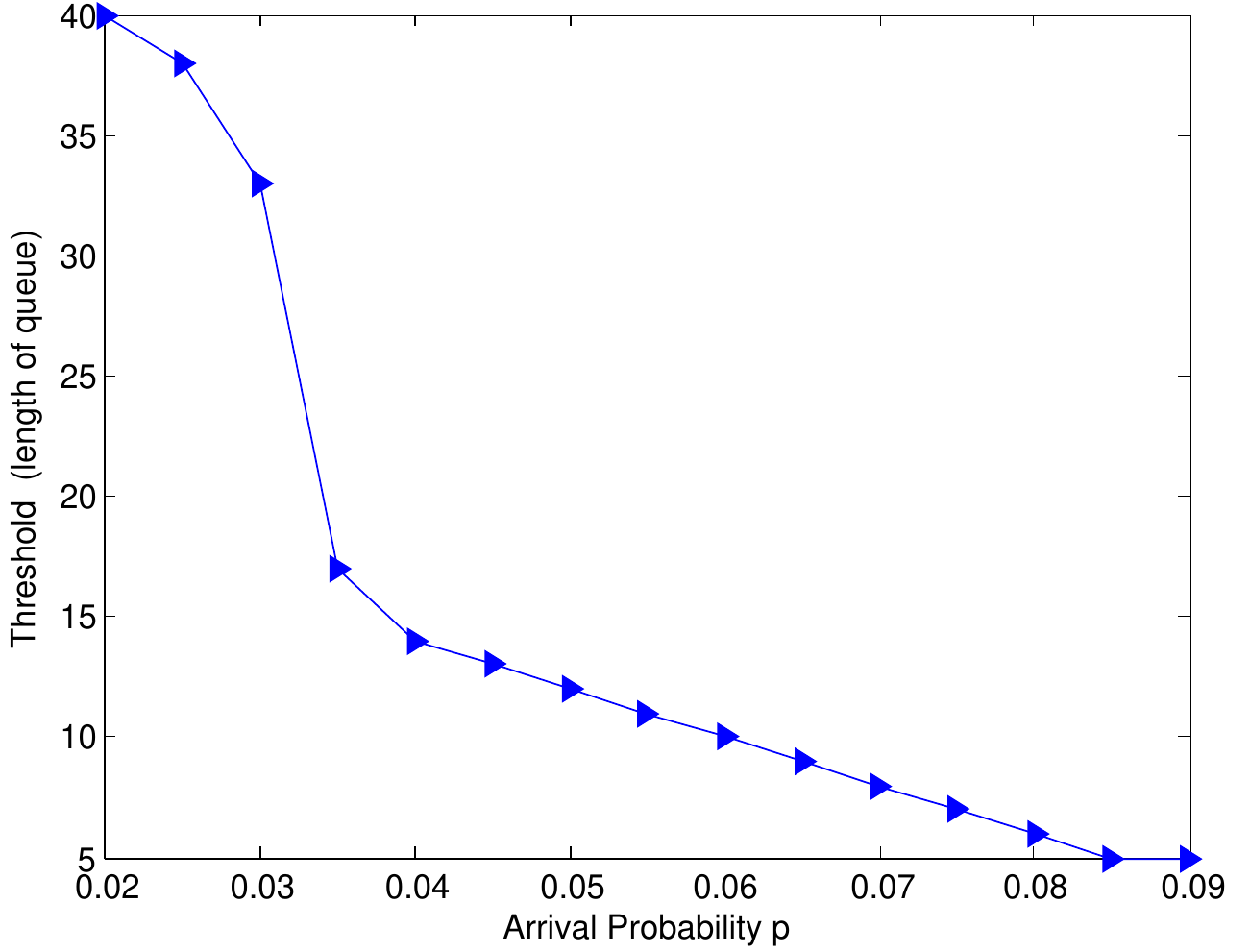}
    \caption{Threshold vs. arrival probability for MDP policy.}
    \label{Fig_sys_MDPthreshold_p}
\end{figure}

\begin{figure}[!t]
    \centering
    \includegraphics[width=0.35\textwidth]{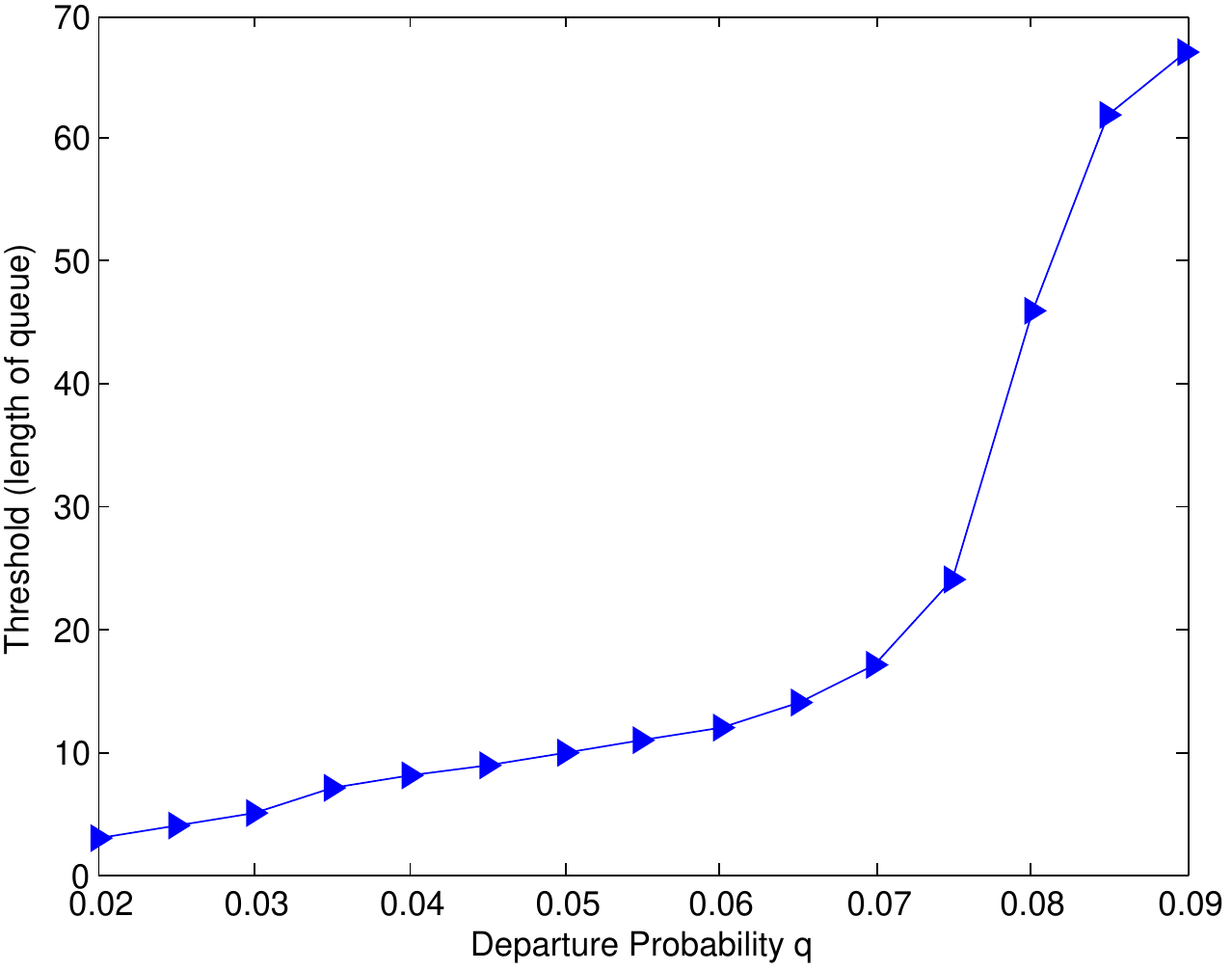}
    \caption{Threshold vs. departure probability for MDP policy.}
    \label{Fig_sys_MDPthreshold_q}
\end{figure}

In our simulation results, Fig. \ref{Fig_sys_MDPthreshold_p} and Fig. \ref{Fig_sys_MDPthreshold_q} show that the new arrival packets will be transferred to core network rather than breakout backhaul when the number of packets in the queue of backhaul reaches a certain value, which also prove the necessity of our strategy analysis. What's more, it can be seen that a higher arrival probability $q$ leads to a smaller threshold value, and a higher departure probability $p$ leads to a bigger threshold value. In fact, the  higher arrival probability or the lower departure probability of packets, the more likely packets wait in line, thus packets will be transferred to core network in a smaller threshold value in this case, and vice verse.


\section{Implementation of Local Breakout Framework}
\label{section5}

We implement a hardware test-bed for the proposed framework of local breakout,  which is realized via programmable software on General Processing Platforms (GPP) on a software defined radio platform.

\begin{figure}[!t]
    \centering
    \includegraphics[width=0.4\textwidth]{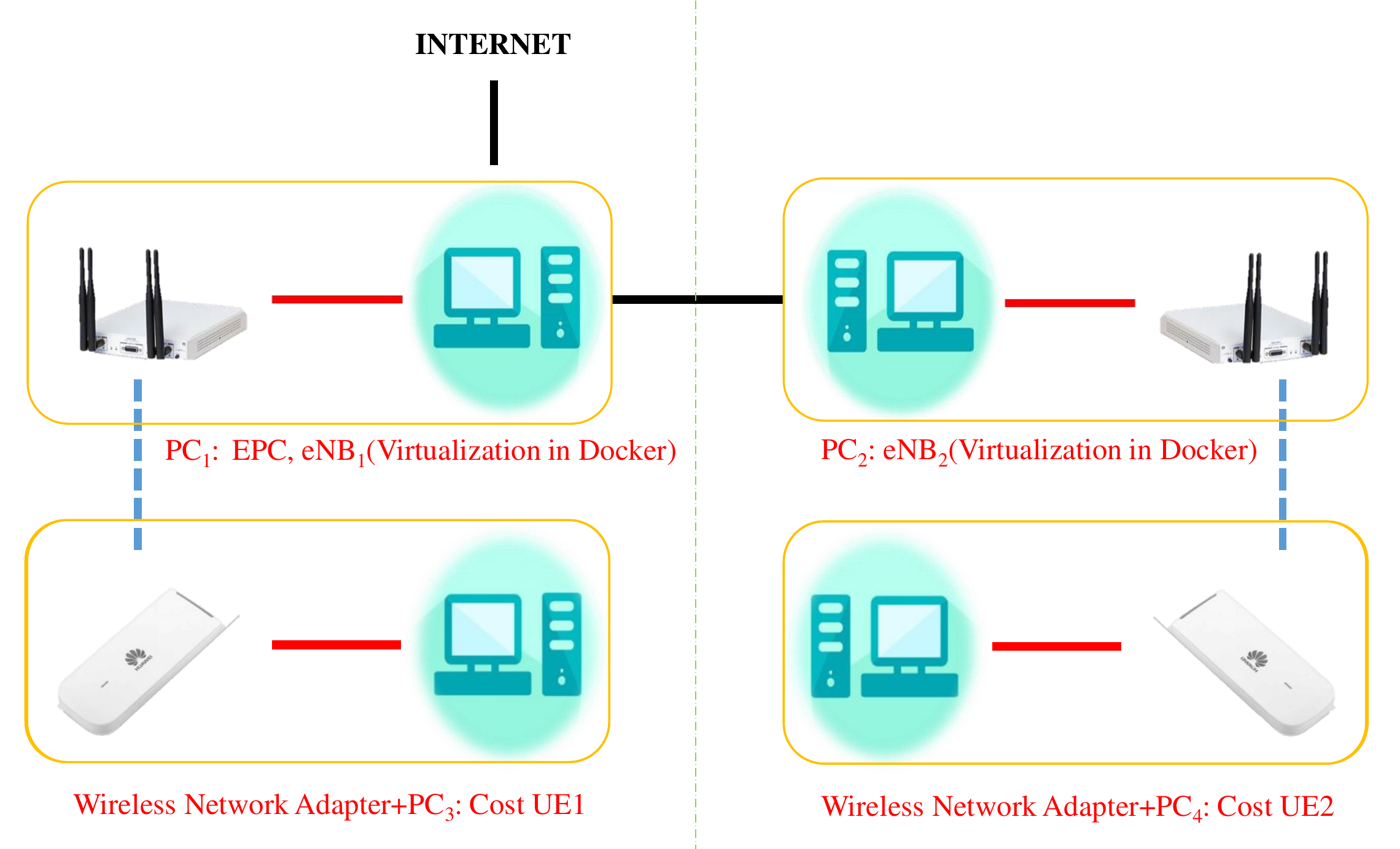}
    \caption{Implementation framework.}
    \label{Fig_sys_implement}
\end{figure}

Based on the source code of the OAI, which is  an open-source ecosystem for the core (EPC) and access-network (EUTRAN) protocols of 3GPP cellular systems with the possibility of interoperating with closed-source equipment in either portion of the network, we build a software defined LTE network on two desk-top PCs with the Ubuntu $14.04$ operating system of version $3.19$ low-latency kernel. The one with i7-3200 CPU serves as the base station and the other with i7-3200  CPU runs the base station and core network instances, and the two PCs are connected by Ethernet. USRP B$210$s from Ettus Research are used as the radio front end, which is connected to the host PC. And also, we use wireless network adapter, connected to PC, as COST UE, and COST UE attached to base station can access internet, as shown in  Fig. \ref{Fig_sys_implement}.  We also introduce the virtualization technology to realize more efficient allocation of physical resources by implementing OAI base stations in Docker container \cite{merkel2014docker}. Our testbed work presents a local breakout scheme within single base station, as well as between adjacent base stations through the backhaul.

% However, the source code of OAI can not support a core network attached with multiple base stations because of the same parameters in different base stations, which leads to the unsuccessful registration in MME which is a unit of the core network. Hence, we modified the source code and adopt different carrier frequencies in different base stations. In our implementation, we build two base stations and the carrier frequencies are $2.66$~GHz and $2.68$~GHz, respectively.%
 In the test we monitor the base stations' activities when mobile user sends out IP packets to another user who is attached to the same base station or the adjacent base station by backhaul. At first we measure the end to end delay, the results show that delay without passing through core network is about $30$~ms, and delay passing through core network is slightly higher, but the gap is not large. This can be interpreted as follows: software defined radio introduces additional delays, leading to the fact that the end to end delay through the backhaul is not very small. Additionally, the core network load is small in our tests, and the core network runs in the same PC with the base station, the additional delays by passing through core network is very small ,which leads to the small difference in the two delays.
 
\begin{figure}[!t]
    \centering
    \includegraphics[width=0.3\textwidth]{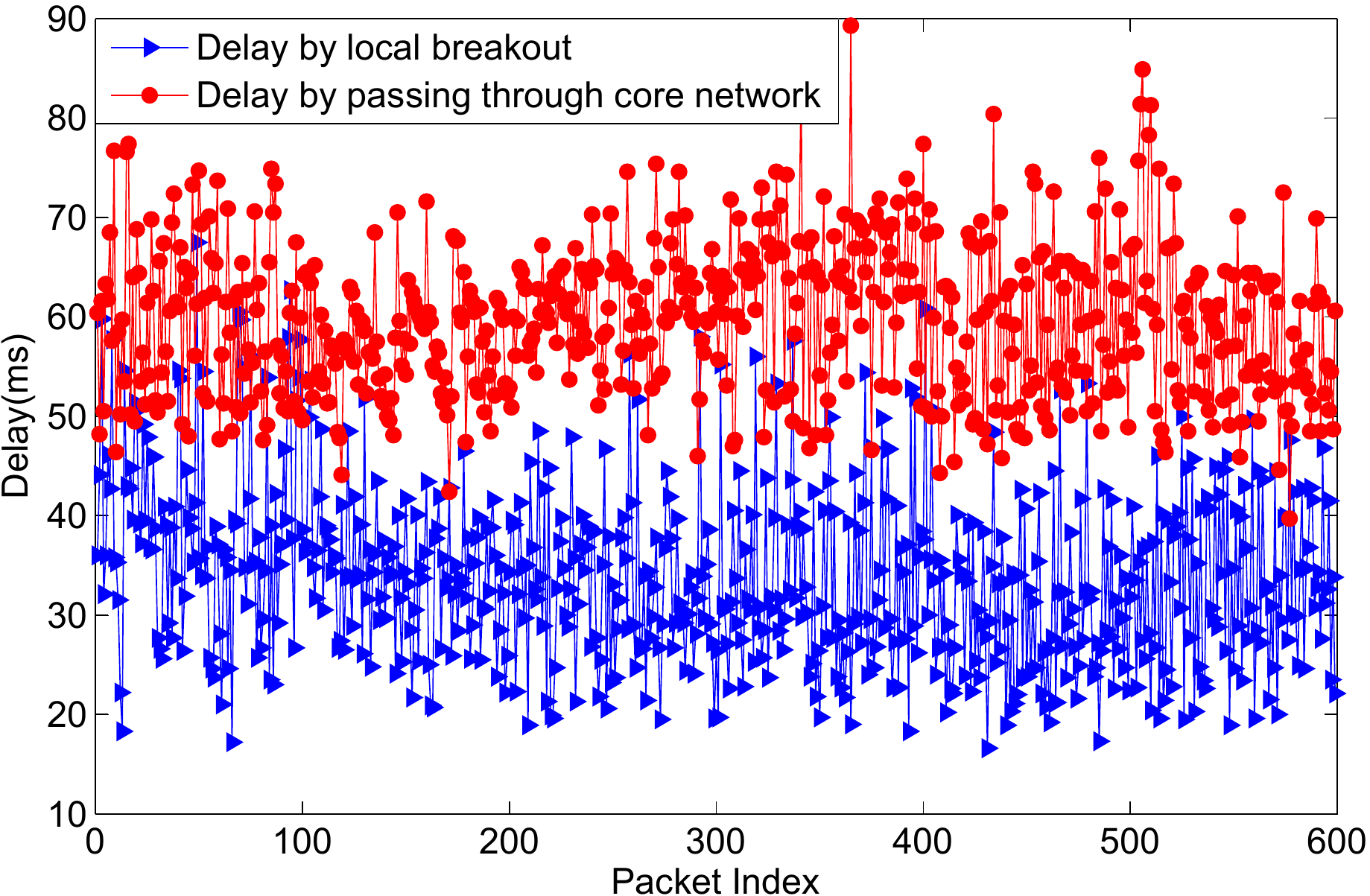}
    \caption{End to end delay of packets by local breakout vs. passing through the core network.}
    \label{Fig_sys_delay}
\end{figure}

Afterwards we deploy the core network in cloud, Fig. \ref{Fig_sys_delay} shows the results that the local breakout framework results in an about $60\%$ delay improvement compared with delay passing through core network.The experiment results validate our new framework of local breakout.
\section{Conclusion}
\label{section6}
In this paper, we propose a local breakout framework between neighbouring base stations to deliver low e2e delay packets with local significance. We also present the validity and practicability of the framework. The problem of designing an efficient packet path choice policy in local breakout framework when the local traffic load is high is investigated. An online MDP based optimum scheduling strategy is proposed. For verification, an OAI local breakout test-bed is implemented, and the virtualization concept is applied to the base stations. The experiment results show that proposed local breakout framework is viable and the e2e delay of packets can be reduced by about $60\%$ through the local link compared to which through the core network.

\section*{Acknowledgment}
This work is sponsored in part by the Nature Science Foundation of China (No. 61701275, No. 91638204, No. 61571265, No. 61621091), the China Postdoctoral Science Foundation, and Intel Collaborative Research Institute for Mobile Networking and Computing.
%%
%\IEEEtriggeratref{7}
\bibliographystyle{ieeetr}
\bibliography{local_BO}

\begin{thebibliography}{10}

\bibitem{index520862global}
C.~V.~N. Index, ``Global mobile data traffic forecast update, 2016--2021 white
  paper,'' {\em URL: http://www. cisco.
  com/c/en/us/solutions/collateral/service-provider/visualnetworking-index-vni/mobile-white-paper-c11-520862.
  html}, Feb 2017.

\bibitem{Andrews14}
J.~Andrews, S.~Buzzi, W.~Choi, S.~Hanly, A.~Lozano, A.~Soong, and J.~Zhang,
  ``What will 5{G} be?,'' {\em IEEE J. Sel. Areas Commun.}, vol.~32,
  pp.~1065--1082, Jun 2014.

\bibitem{boc14}
F.~Boccardi, R.~W. Heath, A.~Lozano, T.~L. Marzetta, and P.~Popovski, ``Five
  disruptive technology directions for {5G},'' {\em IEEE Commun. Maga.},
  vol.~52, pp.~74--80, Feb. 2014.

\bibitem{zheng15}
K.~Zheng, Q.~Zheng, P.~Chatzimisios, W.~Xiang, and Y.~Zhou, ``Heterogeneous
  vehicular networking: A survey on architecture, challenges, and solutions,''
  {\em IEEE Commun. Surveys Tutorials}, vol.~17, pp.~2377--2396, Fourthquarter
  2015.

\bibitem{wang2014cache}
X.~Wang, M.~Chen, T.~Taleb, A.~Ksentini, and V.~Leung, ``Cache in the air:
  exploiting content caching and delivery techniques for 5g systems,'' {\em
  IEEE Communications Magazine}, vol.~52, no.~2, pp.~131--139, 2014.

\bibitem{local2011access}
I.~Local, ``Access and selected {IP} traffic offload ({LIPA-SIPTO}),'' 2011.

\bibitem{cartmell2013local}
J.~Cartmell, J.~McNally, and B.~Balazinski, ``Local selected ip traffic offload
  reducing traffic congestion within the mobile core network,'' in {\em
  Consumer Communications and Networking Conference (CCNC), 2013 IEEE},
  pp.~809--812, IEEE, 2013.

\bibitem{yang2013solving}
M.~J. Yang, S.~Y. Lim, H.~J. Park, and N.~H. Park, ``Solving the data overload:
  Device-to-device bearer control architecture for cellular data offloading,''
  {\em IEEE Vehicular Technology Magazine}, vol.~8, no.~1, pp.~31--39, 2013.

\bibitem{zhang2016mobile}
J.~Zhang, W.~Xie, F.~Yang, and Q.~Bi, ``Mobile edge computing and field trial
  results for {5G} low latency scenario,'' {\em China Communications}, vol.~13,
  no.~Supplement2, pp.~174--182, 2016.

\bibitem{lee2016local}
S.-Q. Lee and J.-u. Kim, ``Local breakout of mobile access network traffic by
  mobile edge computing,'' in {\em Information and Communication Technology
  Convergence (ICTC), 2016 International Conference on}, pp.~741--743, IEEE,
  2016.

\bibitem{samdanis2012traffic}
K.~Samdanis, T.~Taleb, and S.~Schmid, ``Traffic offload enhancements for
  {eUTRAN},'' {\em IEEE Communications Surveys \& Tutorials}, vol.~14, no.~3,
  pp.~884--896, 2012.

\bibitem{ma2011traffic}
L.~Ma and W.~Li, ``Traffic offload mechanism in {EPC} based on bearer type,''
  in {\em Wireless Communications, Networking and Mobile Computing (WiCOM),
  2011 7th International Conference on}, pp.~1--4, IEEE, 2011.

\bibitem{katanekwa2013mobile}
N.~Katanekwa and N.~Ventura, ``Mobile content distribution and selective
  traffic offload in the {3GPP} evolved packet system ({EPS}),'' in {\em
  Information Networking (ICOIN), 2013 International Conference on},
  pp.~119--124, IEEE, 2013.

\bibitem{lee2014novel}
S.-Q. Lee, H.-R. Cheon, S.-H. Kang, and J.-H. Kim, ``Novel {LIPA/SIPTO}
  offloading algorithm according to the network utilization and offloading
  preference,'' in {\em Information and Communication Technology Convergence
  (ICTC), 2014 International Conference on}, pp.~314--318, IEEE, 2014.

\bibitem{bert95}
D.~P. Bertsekas, D.~P. Bertsekas, D.~P. Bertsekas, and D.~P. Bertsekas, {\em
  Dynamic programming and optimal control}, vol.~1.
\newblock Athena Scientific Belmont, MA, 1995.

\bibitem{nikaein2014openairinterface}
N.~Nikaein, M.~K. Marina, S.~Manickam, A.~Dawson, R.~Knopp, and C.~Bonnet,
  ``Openairinterface: A flexible platform for 5g research,'' {\em ACM SIGCOMM
  Computer Communication Review}, vol.~44, no.~5, pp.~33--38, 2014.

\bibitem{jiang_icc17}
Z.~Jiang, S.~Zhou, and Z.~Niu, ``Antenna-beam spatial transformation in {C-RAN}
  with large antenna arrays,'' in {\em IEEE International Conference on
  Communications (ICC) Workshops}, May 2017.

\bibitem{4}
Z.~Jiang, S.~Zhou, and Z.~Niu, ``Dynamic channel acquisition in {MU-MIMO},''
  {\em IEEE Trans. Commun.}, vol.~62, pp.~4336--4348, Dec. 2014.

\bibitem{hooghiemstra2017delay}
G.~Hooghiemstra and P.~Van~Mieghem, {\em Delay distributions on fixed internet
  paths}.
\newblock Delft University of Technology, 2017.

\bibitem{merkel2014docker}
D.~Merkel, ``Docker: lightweight linux containers for consistent development
  and deployment,'' {\em Linux Journal}, vol.~2014, no.~239, p.~2, 2014.

\end{thebibliography}

\end{document}